\documentclass[aps, prb, amsmath,amsfonts]{revtex4}
\usepackage{amsmath}
\usepackage{amssymb}
\usepackage{bm}
\usepackage{graphicx}
\usepackage{times}

\begin{document}

\title{Switching from visibility to invisibility via Fano resonances:\\
theory and experiment}

\author{Mikhail~V.~Rybin${}^{1,2}$}
\email{m.rybin@mail.ioffe.ru}
\author{Dmitry~S.~Filonov${}^{2}$}
\author{Pavel~A.~Belov${}^{2}$}
\author{Yuri~S.~Kivshar${}^{2,3}$}
\author{Mikhail~F.~Limonov${}^{1,2}$}

\affiliation{$^1$ Ioffe Physical-Technical Institute, St.~Petersburg 194021, Russia\\
$^2$ University ITMO, St.~Petersburg 197101, Russia\\
$^3$Nonlinear Physics Center,~Research School of Physics and Engineering,~Australian National University,~Canberra ACT 0200, Australia
}

\begin{abstract}
Subwavelength structures demonstrate many unusual optical properties which can be employed for engineering functional metadevices, as well as scattering of light and invisibility cloaking. Here we demonstrate that the suppression of light scattering for any direction of observation can be achieved for an uniform dielectric object with high refractive index, 
in a sharp contrast to the cloaking with multilayered plasmonic structures suggested previously.
Our finding is based on the novel physics of cascades of Fano resonances observed in the Mie scattering from a
homogeneous dielectric rod. We observe this effect experimentally at microwaves by employing high temperature-dependent dielectric permittivity of a glass cylinder with heated water. Our results open a new avenue in analyzing
the optical response of hight-index dielectric nanoparticles and the physics of cloaking.
\end{abstract}

\date{\today}


\maketitle


\section*{INTRODUCTION}

In the past decade, the study of cloaking and invisibility has attracted a lot of attention. Several approaches
for achieving the cloaking regime have been proposed  on the basis of metamaterials~\cite{zheludev2012metamaterials},
and they employ transformation optics~\cite{leonhardt2006optical,pendry2006controlling,schurig2006metamaterial,cai2007optical},
cancelation of light scattering~\cite{ruan2009temporal,edwards2009experimental,chen2012invisibility},
nonlinear response of multi-shell structures~\cite{zharova2012nonlinear}, as well as the use of
multilayered plasmonic particles~\cite{monticone2013multilayered}, graphene~\cite{chen2011atomically},
and magneto-optical effect with an external magnetic field~\cite{kort2013tuning}. All those approaches
require to employ specially designed 'covering shells' with engineered parameters making difficult
a practical realization of many invisibility concepts~\cite{chen2012invisibility}.

Here we suggest a novel approach for the realization of tunable invisibility cloaking at all
angles of observation that allows a direct switching from visibility to invisibility regimes
and back. Our approach is based on the cancelation of scattering from a homogeneous high-index
dielectric  object \emph{without additional coating layers}. The main idea of our approach is based
of the properties of the characteristic lineshape of the Fano resonance \cite{g704}, and
the cloaking effect is achieved through the resonant interference between a broad background
scattering and narrow Mie resonances realized through the cascades of Fano resonances~\cite{rybin2013mieOE}.
We demonstrate this novel cloaking  effect and its tunability experimentally through a heating-induced change
of dielectric permittivity of a glass tube filled with water.

\section*{RESULTS}

\subsection*{General concept and formalism}

The well-known Mie scattering is described by analytical solutions of
the Maxwell equations for elastic scattering of electromagnetic
waves by a sphere~\cite{hulst1957light,g120,stratton2007electromagnetic}.
If the sphere diameter is comparable with the wavelength of incident light
$2r\sqrt{\varepsilon } /\lambda \approx 1$, the Mie scattering will be driven
by resonances in the dielectric sphere. This leads to the emission of electromagnetic waves by the
particle and interference between the nonresonant scattering from the particle and scattering by narrow Mie modes.
If a spectrally narrow Mie band interacts constructively or destructively with a broad radiation
spectrum, we can expect a Fano-type resonance~\cite{g704},
a resonant wave phenomenon well-known across many different branches of physics~\cite{g901}.
Some characteristic examples of Fano resonance are found in the studies of magnetization~\cite{g709} and electronic polarization properties~\cite{g723}, semiconductor optics~\cite{g707,g705}, electron-phonon coupling in superconductors~\cite{g708,g706,limonov2002superconductivity}, and scattering in
photonic crystals~\cite{g020,g030} and plasmonic
nanostructures~\cite{tribelsky2008light,g908,francescato2012plasmonic}.

\begin{figure}[!t]
\includegraphics{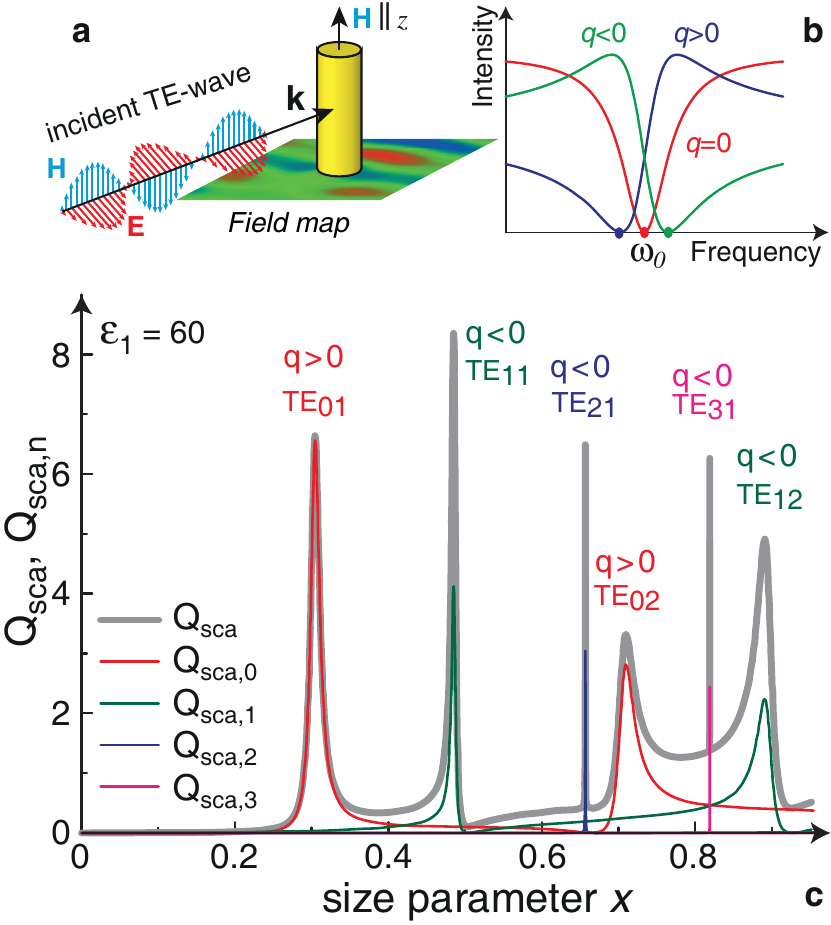}
\caption{
\textbf{Mie scattering from a dielectric rod.}
\textbf{a} Schematic of the scattering geometry for TE polarization.
\textbf{b} Fano lineshapes depending on the sign and value of the Fano parameter
$q$. 
\textbf{c} Spectra of the Mie scattering efficiency
$Q_{\rm sca,n}$ for dipole and multipole modes TE${}_{nk}$, and
for the total scattering efficiency $Q_{\rm sca}$ of a single
dielectric circular rod with $\varepsilon _{1} =60$ and $\varepsilon _{2}
=1$. The sign of the corresponding values of the Fano parameter
$q$ is shown for each Mie resonance. The size parameter $x=r\,\omega/c$.
}
\label{fig:MieSacttering}
\end{figure}

Fano resonance is observed when the wave scattering process
can reach the same final state via two different paths~\cite{g901}.
The first scattering path corresponds to the formation of a narrow
resonant band, where the wave phase changes by $\approx \pi $. The
second scattering path corresponds to a broad background, where the
wave phase and amplitude are nearly constant in the spectrum range of
interest. The resonant band can be described as a complex Lorentz
function $L(\Omega )=(\Omega +i)^{-1} $, where $\Omega =(\omega
-\omega _{0} )/(\Gamma /2)$, while $\omega _{0} $, and $\Gamma $
correspond to the position and the width of the band. The continuum is
defined as $B\exp (i\varphi _{B} )$. The resulting wave takes a form
$A\exp \left(i\varphi _{A} \right)/\left(\Omega +i\right)+B\exp
\left(i\varphi _{B} \right)$ where $A(\omega )$, $B(\omega )$,
$\varphi _{A} (\omega )$ and $\varphi _{B} (\omega )$ are real
functions, frequency dependence of which can be neglected compared to
the Lorentz function. If there is no parts of background avoiding the
above interaction, following Fano relation takes the form
\begin{equation}
I(\omega )=\frac{(q+\Omega )^{2} }{1+\Omega ^{2} } \sin ^{2} [\Delta
(\omega) ]
,\label{eq:Fano}    
\end{equation}
where $q=\cot \Delta $ is the Fano asymmetry parameter, $\sin ^{2}
[\Delta (\omega )]$ represents a background produced by a plane wave,
$\Delta (\omega )=\varphi _{A} (\omega )-\varphi _{B} (\omega )$ is
the phase difference between the narrow resonant and continuum states,
$B(\omega )=\sin [\Delta (\omega )]$ \cite{SMPDF}.
At $q=0$, instead of the conventional Mie peak, the resonant Mie dip with symmetrical Lorentzian shape $I(\omega ){\rm \sim }\Omega ^{2} /\left(1+\Omega^{2} \right)$ is observed with the scattering intensity vanishing at the eigenfrequency $\omega _{0} $.
Destructive interference results in complete suppression of the scattering intensity at a given frequency~\cite{g704}.
When the Fano parameter $q$ deviates from zero, the Fano lineshape becomes asymmetric, and the zero-intensity frequency becomes shifted away from the resonance $\omega _{0}$. It is important to emphasize that \emph{at any finite value of $q$ there exists zero-intensity frequency }$\omega _{{\rm zero}}=\omega _{0} -q\Gamma /2$, see Eq.~(\ref{eq:Fano}). Note that $\omega _{{\rm zero}}<\omega _{0} $ at $q>0$ and $\omega _{{\rm zero}} >\omega _{0} $ at $q<0$ that is clearly seen in Fig.~\ref{fig:MieSacttering}\textbf{b}. We employ this property
of the Fano lineshape in order to realize the invisibility cloaking for a high-index dielectric rod in free space.

\subsection*{Fano resonances in the Mie scattering}

\begin{figure*}[t!]
\includegraphics{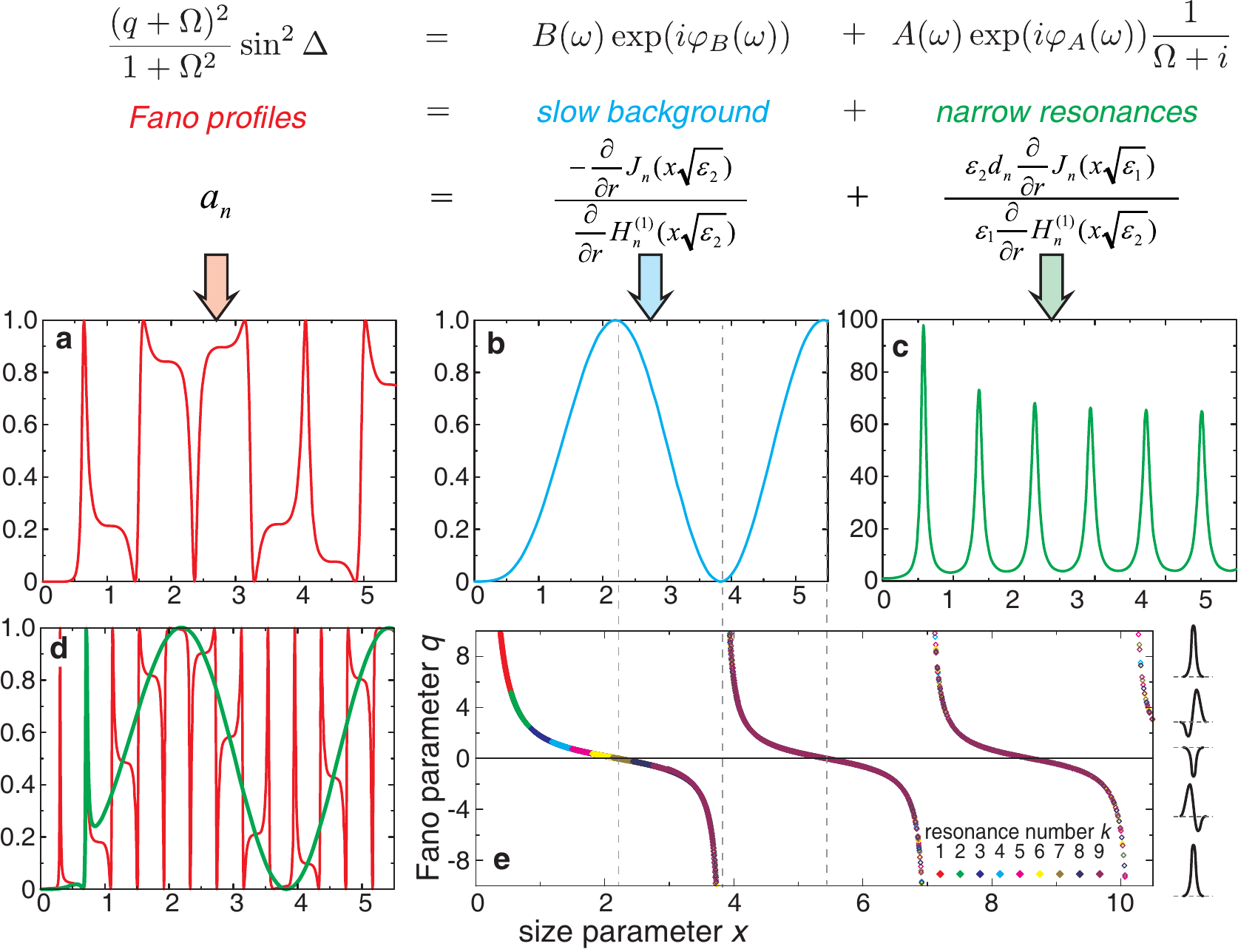}
\caption{
\textbf{Fano resonances for the Mie resonant modes.} 
\textbf{a} Spectra of squared modules of the Lorenz-Mie coefficient $|a_0|^2$ describing magnetic field outside a circular rod with $\varepsilon_1 = 16$. 
\textbf{b} The background spectrum.
\textbf{c} Spectra of squared modules of the Lorenz-Mie coefficient $|d_0|^2$ describing magnetic field inside a rod with $\varepsilon_1 = 16$.
\textbf{d} Squared Lorenz-Mie coefficient $|a_0|^2$ for a rod with $\varepsilon_1 = 60$ and an example of the Fano fitting of the TE$_{02}$, mode (green curve).
\textbf{e} Calculated dependence of the Fano parameter $q$ on the size parameter $x = r\omega /c = 2\pi r/\lambda $ for the dipole modes TE$_{0k}$ ($1 \leqslant k \leqslant 9$) of a circular rod ($1 \leqslant \varepsilon_1 \leqslant 150$) embedded in air ($\varepsilon_2 = 1$). Four characteristic Fano lineshapes for selected values of $q$ are shown on the right.
 \textbf{Top} The equation expressed the Fano resonance condition and the Maxwell boundary condition. $J_n(\zeta)$ and $H^{(1)}_n (\zeta)$ are the Bessel and Hankel functions.
}
\label{fig:FanoLM}
\end{figure*}

\begin{figure*}[ht]
\includegraphics[scale=1.3]{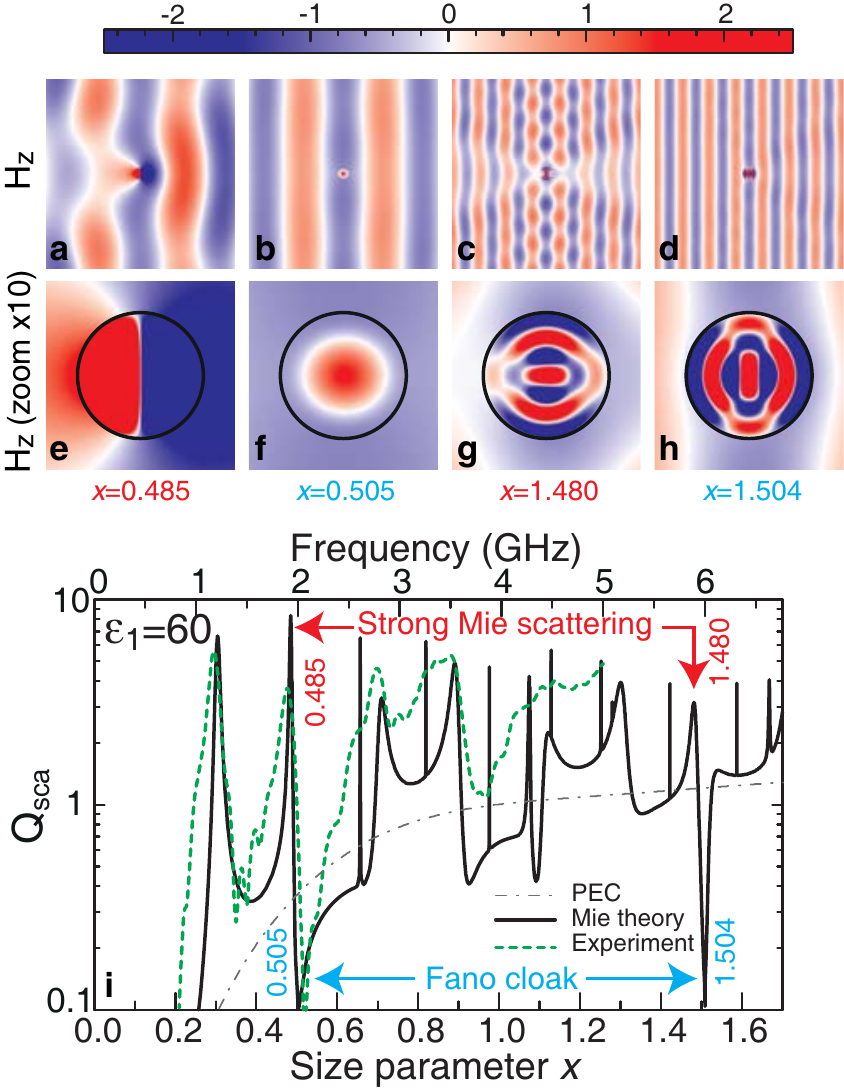}
\caption{
\textbf{Magnetic field map and scattering efficiency spectra for TE polarization.}
\textbf{a}-\textbf{h} Results of numerical calculations for the Mie scattering by a single
cylinder ($\varepsilon _{1} =60$) embedded in air ($\varepsilon _{2}
=1$). \textbf{a}-\textbf{h} $H_{z} $ component of the TE polarized electromagnetic
field \textbf{a}-\textbf{d} around cylinder and \textbf{e}-\textbf{h} inside cylinder. Shown are:
strong Mie scattering (uncloaked) regimes at \textbf{a},\textbf{e} $x=0.485$ and \textbf{c},\textbf{g}
$x=1.48$, as well as Fano cloaking regimes at $x_{{\rm cloak}} =0.505$
\textbf{b},\textbf{f} and $x_{{\rm cloak}} =1.504$ \textbf{d},\textbf{h}.
\textbf{i} Spectral dependence of the scattering efficiency $Q_{sca} $ (black curve), scattering efficiency
of the perfect metallic conductor (gray dash-and-dot curve), and the
experimentally measured scattering efficiency of water's cylinder in
free space (green dashed curve).
}
\label{fig:FieldsQsca}
\end{figure*}

Our study outlined below is limited to the case of an infinitely long dielectric
circular rod. However, the suggested concept and subsequent analysis are rather general, and they
can be applied to other types of ``bodies of revolution''. The case of an infinite rod,
referred to as the Lorenz-Mie theory, corresponds to a two-dimensional problem of scattering
in the plane normal to the symmetry axis $z$, and it involves cylindrical rather then spherical
functions in the infinite series of the analytical Mie
solution~\cite{g120,hulst1957light,stratton2007electromagnetic}.


We consider the Mie scattering by a single homogeneous
infinite circular rod with the radius $r$ and the purely real dielectric permittivity
$\varepsilon _{1} $ embedded in the transparent and homogeneous
surrounding medium with the dielectric permittivity of $\varepsilon
_{2} = 1 $. The Mie scattering by a cylinder can be expanded into
orthogonal electromagnetic dipolar and multipolar terms, with
cylindrical Lorenz-Mie coefficients $a_{n} $ and $b_{n} $ \cite{g120}. For the
TE-polarization considered here, the scattered fields are defined by
coefficients of only one type $a_{n} $, while $b_{n} $ are equal to
zero. In addition, we introduce new coefficients $d_{n} $ to
characterize the field inside the cylinder; more details are given in Methods. The resonances are denoted
as TE${}_{nk} $ where $n$ is multipole order, and $k$ is resonance
number ($n$ is integer and $k$ is positive integer).

Figure~\ref{fig:MieSacttering}\textbf{c} presents the total Mie scattering efficiency $Q_{\rm sca}
=\frac{2}{x} \sum _{n=-\infty }^{\infty } \left|a_{n} \right|^{2} $
and
the spectral dependence of the Mie
scattering efficiency of individual modes $Q_{\rm sca,n} =\frac{2}{x}
\left|a_{n} \right|^{2} $ in the low-frequency part of the spectrum at
$\varepsilon _{1} = 60$ where we observe the modes TE$_{0k} $, TE$_{1k}
$, TE$_{2k} $ and TE$_{3k} $. Intensity of the multipole modes
TE$_{4k} $, TE$_{5k} $,  ($n>3$) differs from zero for higher
frequencies for $x>1$.
%
Spectra of the Mie scattering presented in Fig.~\ref{fig:MieSacttering}\textbf{c} demonstrate
different possible lineshapes that can be described by the Fano formula~(\ref{eq:Fano}).

We show in Fig.~\ref{fig:FanoLM} and in Methods how the spectrum of the Mie scattering can be presented in
the form of an infinite series of the Fano profiles. We consider the
Maxwell boundary condition for the tangential components of $E$ field
at the cylinder surface for the TE polarization \cite{SMPDF}.
The interference of the background and resonant scattered fields creates a cascade of
Fano profiles (see Fig.~\ref{fig:MieSacttering}\textbf{c} or \ref{fig:FanoLM}). In contrast to the spectra of $\left|d_{0}
\right|^{2} $ describing magnetic field inside the rod, the scattering spectra
outside the rod $\left|a_{0} \right|^{2} $ demonstrate
asymmetric profiles with either sharp increase or drop of $\left|a_{0}
\right|^{2} $ values at the resonance frequencies of the cylinder
eigenmodes.
To demonstrate that we really have the Fano resonance, we calculated the spectral dependence
of the Fano parameter $q$  for the dipole mode TE$_{0k}$ using the profiles of 2700 resonances (for $1\leqslant k \leqslant 9$ and for $1\leqslant \varepsilon_1 \leqslant 150$ with the step of $\Delta\varepsilon = 0.5$) in the broad spectral range at $\varepsilon_2 = 1$. Figure~\ref{fig:FanoLM}\textbf{d} shows an example of such a fitting for TE$_{02}$ mode at $\varepsilon_1 = 60$ with Fano parameter $q=3.55$ at $x=0.71$. Finally we obtained an important result: all 2700 values of the Fano parameter $q$ forms a general dependence $q(x)$ (Fig.~\ref{fig:FanoLM}\textbf{e}). As follows from the fitting results, the roots of the background function define the points of infinity $q(x) \to \pm\infty$ while the maxima of the background define the roots of $q(x)$ (see common vertical dashed lines in Figs.~\ref{fig:FanoLM}\textbf{b}~and~\ref{fig:FanoLM}\textbf{e}). The Fano parameter demonstrates characteristic cotangent-type dependence that is the key feature of the Fano approach demonstrating that the Fano lineshape depends only on the position of the resonance on the frequency scale with respect to the maximum (or minimum) of the background. Each Mie resonance is phase shifted with respect to the maximum of the sin - type background by phase $\Delta(\omega)$. In particular, Mie resonances positioned close to the background maxima (at $x \approx 2.2$, $5.4$ etc) exhibit symmetric Lorentzian-type dips while Mie resonances positioned close to the background minima (at $x =0$, $\approx 3.8$ etc) exhibit symmetric Lorentzian-type peaks. On the both sides of the maxima mirror-like asymmetric resonance profiles are observed. This law is valid independently on the rod permittivity $\varepsilon_1$.



Numerically calculated structure of the magnetic field in the regimes of Fano cloaking and strong Mie scattering
are shown in Fig.~\ref{fig:FieldsQsca}\textbf{a}-\textbf{h}. The supplementary video \cite{SMAVI} demonstrates the Mie scattering by a dielectric cylinder in the spectral range $0\leqslant x \leqslant 2$. We observe a practically complete suppression
of scattering at frequencies corresponding to the dips in the function $Q_{\rm sca} (x)$ (Fig.~\ref{fig:FieldsQsca}\textbf{i}. It means that the incident TE-polarized light passes the cylinder without scattering making the cylinder invisible from {\em any angle of observation}.
To analyze the invisibility dynamics, we calculated the dip intensity as a function of dielectric constant $\varepsilon_1$ for the lowest dip in the frequency scale. For a reference scattering intensity we choose standard scattering efficiency for cylinder made from perfect conducting metal (PEC). The calculations shows that the scattering efficiencies of dielectric cylinder and PEC coincide when cylinder permittivity is about 10. At $\varepsilon_1>10$ the invisibility dynamics appears for $x$ around $0.505$.

\subsection*{Experimental verification of tunable invisibility}

\begin{figure}[t]
\includegraphics{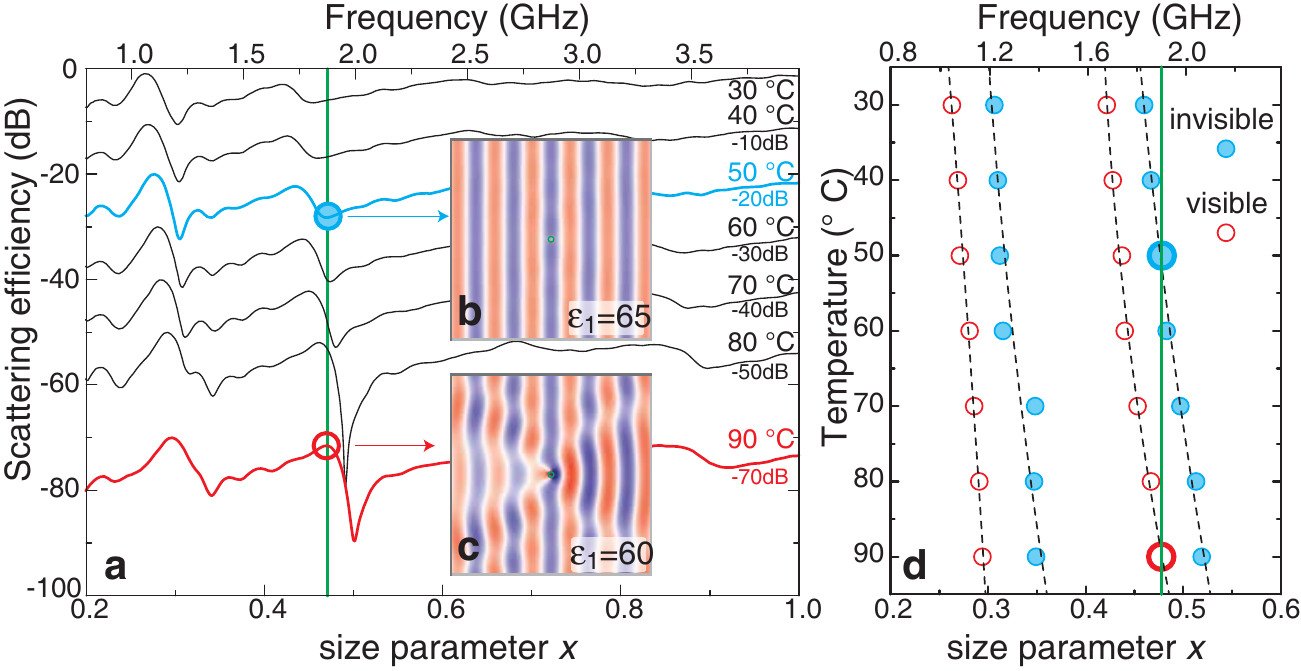}
\caption{
\textbf{Experimental demonstration of tunable invisibility.}
\textbf{a} Measured temperature dependence of the scattering efficiency of a glass tube
filled with water. Curves are shifted vertically by the values marked on the plot. Inserts \textbf{b},\textbf{c} show the calculated magnetic fields at the frequency
1.9 GHz in the regimes of the Fano invisibility and strong Mie scattering, respectively.
\textbf{d} Measured positions of dips in the scattering efficiency, corresponding to invisible cylinder, and peaks, corresponding visible cylinder and strong Mie scattering.
}
\label{fig:switching}
\end{figure}

In the case of high-index dielectric materials and weak losses, a typical asymmetric Fano profile has a local maximum and a local minimum located close to each other, as shown in Fig.~\ref{fig:FieldsQsca}\textbf{i}.
The existence of the first strong dip at $\omega _{{\rm cloak}} =0.505c/r$ ($\varepsilon _{1} =60$) is caused by several reasons: the perfect zero-intensity condition at $\omega _{{\rm zero}} >\omega _{0} $ ($q<0$) for nearby Fano-type TE$_{11}$ mode, narrowness of the neighboring TE$_{21}$ mode and long-distance location to two intense scattering bands TE$_{01}$ and TE$_{02}$ (Fig.~\ref{fig:MieSacttering}\textbf{c}).
 This proximity  can be employed for the demonstration of the switching between viability and  invisibility, tuning the scattering from the uncloaked to cloaked regimes. This can be achieved by modulating the  parameters of an object or by changing  the wavelength of the incoming radiation. To demonstrate the concept of Fano cloaking in experiment, we employ the advantages of strong temperature dependence of the dielectric permittivity of water~\cite{zelsmann1995temperature}.

At microwave frequencies, we use a glass cylinder filled with water that is characterized by dielectric constant
of $\varepsilon=80$ at 20${}^\circ$C and $\varepsilon=50$ at 90${}^\circ$C in the frequency
range from 1 GHz to 6 GHz. A rectangular horn antenna (TRIM 0.75 GHz to 18 GHz; DR) connected to a transmitting port of the vector network analyzer Agilent E8362C is used to approximate a plane-wave excitation.  The water cylinder with radius $12$~mm  and height $42$~cm is placed into the far-field region  of the antenna (on the distance approximately 2.5 m) and the similar horn antenna (TRIM 0.75 GHz to 18 GHz) is employed as a receiver. The scattering efficiency is yielded from the imaginary part of the forward scattering amplitude (due to the optical theorem). The latter is proportional to $E/E_0-1$,
where $E_0$ is measured electric field in the free space, and $E$ is the electric field in the presence
of the cylinder with water.

Figure~\ref{fig:FieldsQsca}\textbf{i} shows strong suppression of scattering (of the order of 20 dB) from a glass tube filled with water and measured in microwave experiment; the data agree well with the theoretical predictions. A strong temperature dependence of the dielectric permittivity of water $\varepsilon _{1}$ leads to a profound shift of the scattering maxima (the uncloaked regime) and minima (the cloaked regime) located in the spectrum near the resonant frequencies of the Mie resonance modes TE$_{01}$ (the spectral region $\omega {\rm \sim 1.25}$GHz) and TE${}_{11}$ ($\omega  \sim 2$GHz), as shown in Fig.~\ref{fig:switching}.

\section*{DISCUSSION}

A proximity of the maxima and minima of the Fano resonance in the frequency scale
allows us to switch a glass tube with water from the visible regime of strong Mie scattering ($T=90^\circ$C) to the regime of strong invisibility ($T=50^\circ$C) \emph{at the same frequency of 1.9 GHz}. Numerical calculations of the  component $H_{z}$ of the electromagnetic field confirm directly the switching effect, and demonstrate both perfect invisibility in Fig.~\ref{fig:switching}\textbf{b} and strongly distorted scattered fields in Fig.~\ref{fig:switching}\textbf{c}.

So, we have presented the first experimental observation of tunable invisibility of a macroscopic object
transformed from visible to invisible states and back, without any coating layers. We have shown that the total intensity of the Mie scattering for waves of a certain polarization vanishes under the condition of the Fano resonance at any angle of observation. Remarkable that high-index dielectric materials are available for different wavelength ranges or can be engineered at will~\cite{hu2013electron}. Our study reveals a novel physics behind the seemingly well-known Mie scattering of light and it may open a novel route towards manipulation and control of electromagnetic waves in all-dielectric nanophotonics.

\section*{Acknowledgements}

We acknowledge fruitful discussions with A.E.~Miroshnichenko, A.N.~Poddubny, Yu.A.~Baloshin, and A.P.~Slobozhanyuk. This work was supported by the Government of the Russian Federation (grant 074-U01), the Ministry of Education and Science of the Russian Federation, Russian Foundation for Basic Research (project no. 13-02-00186), Dynasty Foundation (Russia), and the Australian National University.


\section*{Author Contributions}
M.R. developed a theoretical model and conducted simulations and data analysis. D.F.~performed experimental measurement. M.L, P.B. and Y.K. provided a guidance on the theory, numerical analysis and experiment. All authors discussed the results and contributed to the writing of the manuscript.

\section*{Additional information}
The authors declare no competing financial interests.
Correspondence and requests for materials should be addressed to M.R. (Email: m.rybin@mail.ioffe.ru)

\section*{METHODS}


{\em Fano formula.} First, we consider a classical problem of the Fano resonance that appears as a
result of interference between a narrow (resonant) band and a continuum background
known as the configuration interaction in the physics of quantum phenomena.
The resonant band can be described by a complex Lorentzian function $L(\Omega )=(\Omega +i)^{-1} $, where
$\Omega =(\omega -\omega _{0} )/(\Gamma /2)$, while $\omega _{0} $,
and $\Gamma $ corresponds to the position and width of the narrow frequency band.

We can define the continuum spectrum as $B\exp (i\varphi _{B} )$, so that the resulting
combined wave takes the form
\begin{equation}
A(\omega )\exp (i\varphi _{A} (\omega ))\frac{1}{\Omega +i} +B(\omega
)\exp (i\varphi _{B} (\omega )).
\end{equation} 
Here $A(\omega )$, $B(\omega )$, $\varphi _{A} (\omega )$, and $\varphi
_{B} (\omega )$ are real functions which relative changes in the
frequency range of interest are negligible in comparison with the
Lorentzian function. The intensity of the resulting wave is given by
\begin{equation}
I(\omega )=\left|\frac{A\exp (i\Delta )+B(\Omega +i)}{\Omega +i}
\right|^{2} ,
\label{eq:Intens}
\end{equation} 
where $\Delta (\omega )=\varphi _{A} (\omega )-\varphi _{B} (\omega )$
is the phase difference between the resonant and continuum states. The
resulting extended Fano formula can be written in the form
\begin{equation}
I(\omega )=\left(\frac{(q+\Omega )^{2} }{1+\Omega ^{2} } \eta
+(1-\eta )\right)B^{2} .
\label{eq:FanoExt}
\end{equation} 

In Eq.~(S\ref{eq:FanoExt}), the parameter $q$ is defined as the Fano asymmetry parameter
that characterizes a relative transition strength for the discrete
state vs. continuum set of states. The first term in Eq.~(S\ref{eq:FanoExt})
describes a narrow band, and the additional background spectrum in
the region of the narrow band is presented by the second term. The
background component that does not interfere with the narrow band is
accounted for by the introduction of an interaction coefficient $\eta \in
[0..1]$. Such non-interacting background is observed experimentally
in the light scattering from different physical systems
\cite{g708,limonov2002superconductivity,g706} and photonic crystals
\cite{g020,g026}. Comparing Eq.~(S\ref{eq:Intens})~and~ Eq.~(S\ref{eq:FanoExt}),
 we obtain
\begin{equation}
\left\{\begin{array}{c} { q=\frac{\displaystyle F+2\sin \Delta
+\sqrt{F^{2} +4F\sin \Delta +4} }{\displaystyle 2\cos \Delta } }\\
\,\,\, \\
 {\eta =\frac{\displaystyle 2F\cos ^{2} \Delta }{\displaystyle F+2\sin \Delta +\sqrt{F^{2}
+4F\sin \Delta +4} } } \end{array}\right.
,\label{eq:FanoParams}
\end{equation} 
where $F=A/B$ is the relative intensity of the narrow band and
background component.

For $\eta=1$ and $A=1$, we obtain the Fano lineshape profile,
\begin{equation}
I(\omega )=\frac{(q+\Omega )^{2} }{1+\Omega ^{2} } \sin ^{2} \Delta
,
\label{eq:theFano}
\end{equation} 
where $q=\cot \Delta $ and $B=\sin \Delta $. We notice that the
absolute value of the Fano parameter $q$ is a measure of the relative
strength of the amplitude of the resonant scattering compared to its
nonresonant value.

{\em Mie scattering as a cascade of Fano resonances.}
To demonstrate that the spectrum of Mie scattering can be presented in
the form of an infinite series of the Fano profiles, we consider
elastic scattering of electromagnetic waves by a homogeneous infinite
dielectric rod of the radius $r$ and the purely real dielectric
permittivity $\varepsilon _{1} $. The surrounding medium have the
dielectric permittivity of $\varepsilon _{2} $. The far-field
scattered by a cylindrical rod can be expanded into orthogonal
electromagnetic dipolar and multipolar terms, with cylindrical
Lorenz-Mie coefficients $a_{n} $ and $b_{n} $ \cite{g120}. For the
TE-polarization the scattered fields are defined by coefficients of
only one type $a_{n} $, while $b_{n} $ are equal to zero. The Maxwell
boundary conditions for the tangential components of H and E fields at
the rod's surface for the TE polarization can be presented as
\begin{equation}
\left\{
\begin{array}
{l} {E_{n} J_{n} (x\sqrt{\varepsilon _{2} }
)+A_{n} H_{n}^{(1)} (x\sqrt{\varepsilon _{2} } )=D_{n} J_{n}
(x\sqrt{\varepsilon _{1} } )} \\
{\varepsilon _{1} E_{n} \frac{\partial }{\partial r}
J_{n} (x\sqrt{\varepsilon _{2} })+\varepsilon _{1}
A_{n} \frac{\partial }{\partial r} H_{n}^{(1)}
(x\sqrt{\varepsilon _{2} } )=\varepsilon _{2} D_{n}
\frac{\partial }{\partial r} J_{n} (x\sqrt{\varepsilon _{1} } )}
\end{array}\right.
\label{eq:boundcond}, 
\end{equation}
where $x=r\omega /c=2 \pi \, r/\lambda $, $E_{n} $, $A_{n} $
and $D_{n} $ are cylindrical harmonic amplitudes of the incident,
scattered and internal magnetic fields, respectively, expressed in
terms of the Bessel $J_{n} (\zeta )$ and Hankel $H_{n}^{(1)} (\zeta )$
functions. The Lorenz-Mie coefficients $a_{n} =A_{n} /E_{n} $ and
coefficients $d_{n} =D_{n} /E_{n} $ for the scattered and internal
magnetic fields can be determined from the system (S\ref{eq:boundcond}) as:
\begin{equation}
a_{n} =\frac{\varepsilon _{2} J_{n} (x\sqrt{\varepsilon _{2} }
)\frac{\partial }{\partial r} J_{n} (x\sqrt{\varepsilon _{1} }
)-\varepsilon _{1} \frac{\partial }{\partial r} J_{n}
(x\sqrt{\varepsilon _{2} } )J_{n} (x\sqrt{\varepsilon _{1} }
)}{\varepsilon _{1} \frac{\partial }{\partial r} H_{n}^{(1)}
(x\sqrt{\varepsilon _{2} } )J_{n} (x\sqrt{\varepsilon _{1} }
)-\varepsilon _{2} H_{n}^{(1)} (x\sqrt{\varepsilon _{2} }
)\frac{\partial }{\partial r} J_{n} (x\sqrt{\varepsilon _{1} } )},
\label{eq:an_ext} 
\end{equation}
\begin{equation}
d_{n} =\frac{J_{n} (x\sqrt{\varepsilon _{2} } )\frac{\partial
}{\partial r} H_{n}^{(1)} (x\sqrt{\varepsilon _{2} } )-H_{n}^{(1)}
(x\sqrt{\varepsilon _{2} } )\frac{\partial }{\partial r} J_{n}
(x\sqrt{\varepsilon _{2} } )}{J_{n} (x\sqrt{\varepsilon _{1} }
)\frac{\partial }{\partial r} H_{n}^{(1)} (x\sqrt{\varepsilon _{2} }
)-\frac{\varepsilon _{2} }{\varepsilon _{1} } H_{n}^{(1)}
(x\sqrt{\varepsilon _{2} } )\frac{\partial }{\partial r} J_{n}
(x\sqrt{\varepsilon _{1} } )}.
\label{eq:dn_int} 
\end{equation}

Figure~\ref{fig:FanoLM} demonstrates that the spectrum of Mie scattering can be
presented in the form of an infinite series of the Fano profiles. We
write simultaneously two relations (i) the expression for the
Lorenz-Mie coefficient $a_{n} $ which defines the scattered field in
accord to (\ref{eq:boundcond}), (ii) analytical expression describing the Fano resonance
at the interference of a narrow resonance (symmetric Lorentzian) with
a slow varying background (plane wave).

We identify two terms of different line-width with the first one
$\left[-
\frac{\partial }{\partial r} J_{n} (x\sqrt{\varepsilon _{2} }
)\right]/ \left[\frac{\partial }{\partial r} H_{n}^{(1)} (x\sqrt{\varepsilon _{2} } ) \right]$
as slowly changed background and second one,
$\left[ \varepsilon _{2} d_{n}
\frac{\partial }{\partial r} J_{n} (x\sqrt{\varepsilon _{1} } ) \right] / \left[\varepsilon _{1} \frac{\partial }{\partial r} H_{n}^{(1)} (x\sqrt{\varepsilon _{2} } ) \right]$
as symmetric Lorentzian, both oscillating with the resonance frequency $\omega$. Therefore we have a Fano resonance between background scattering from the rod and resonant Mie mode within the same cylindrical harmonic (see Fig.~\ref{fig:FanoLM}).
The interference of the incident and scattered fields creates a
complicated near-field pattern, and it may give rise either to strong
enhancement (constructive interference) or strong suppression
(destructive interference) of the electromagnetic field around the
rod.

The similar analysis can be performed in the case of the TM polarized
waves and for any ``body of revolution'' including a sphere.


\end{document}